\renewcommand{\theequation}{\thesection.\arabic{equation}}

\documentstyle[12pt]{article}

\begin{document}

\setlength{\baselineskip}{0.2in}

\begin{titlepage}
\noindent
\hfill{HUPD-9419}\par
\hfill{December 1994}

\vskip 1.0cm
\Large
\centerline{\bf $Q^2$-Evolution of Nucleon's Chiral-Odd Twist-3}
\centerline{\bf Structure Function: $h_L(x,Q^2)$}

\vskip1.5cm

\normalsize
\centerline{Yuji Koike}
\vskip0.2cm
\centerline{\it Department of Physics, Niigata University,
Niigata 950-21,
Japan
\footnote{Present address.
E-mail: koike@NUCL01.sc.niigata-u.ac.jp
}}
\centerline{\it and}
\centerline{\it National
Superconducting Cyclotron Laboratory, Michigan State University}
\centerline{\it East Lansing, MI 48824, USA}

\vskip0.7cm
\centerline{\it and}
\vskip0.7cm

\centerline{Kazuhiro Tanaka}
\vskip0.2cm
\centerline{\it Department of Physics, Hiroshima University,
Higashihiroshima-shi, Hiroshima 724,
Japan\footnote{Present address.
E-mail: tanakak@theo.phys.sci.hiroshima-u.ac.jp}}
\centerline{\it and}
\centerline{\it Research Center for Nuclear Physics,
Osaka University,
Ibaraki-shi, Osaka 567, Japan}

\vskip2.0cm
{\bf Abstract:}
We investigate the $Q^{2}$-evolution of the chiral-odd
spin-dependent
parton distribution $h_{L}(x, Q^{2})$ relevant
for the polarized
Drell-Yan processes.
The results are obtained in the leading logarithmic order
in the framework of the
renormalization group and the standard QCD perturbation theory.
We calculate the anomalous dimension matrix
for the twist-3 operators for $h_{L}$ in the one-loop order.
The operator mixing among the relevant twist-3 operators
including the operators proportional to the QCD equations
of motion
is treated properly in a consistent scheme.
Implications for future
experiments are also discussed.

\vskip0.5cm
\end{titlepage}

\newpage
\section{Introduction}
\setcounter{equation}{0}
\renewcommand{\theequation}{\arabic{section}.\arabic{equation}}

The EMC measurement of the $g_1$-structure function of the
nucleon\,\cite{EMC} has given rise to renewed interest in the
spin effects in high energy processes.
In particular, many authors have been
attracted by the chiral-odd spin-dependent
distribution functions $h_{1,L}$\cite{AM,CPR,JJ,C}.
Because of the chirality, they can not be measured
in the totally inclusive deep inelastic
lepton-nucleon scattering
(except as a quark-mass effect), but reveal themselves in
polarized
hadron-hadron collisions
(Direct photon production, Drell-Yan process, etc.)
and semi-inclusive polarized electron scatterings.
Thus they are expected to open up a new window to explore
hadron structure.

The twist-2 parton distribution $h_1$ was first
addressed by Ralston and Soper more than ten years ago
\cite{RS} and was recently named as
the ``transversity'' distribution in ref. \cite{JJ}.
It has a simple parton model interpretation like $f_1$ and
$g_1$,\footnote{We denote the twist-2 distributions
corresponding to
the structure functions $F_{1}$ ($F_{2}$) and $g_{1}$ by
$f_{1}$ and $g_{1}$ following ref.\cite{JJ}.} and
appears as a leading contribution, for example, in the
Drell-Yan process with the
transversely polarized nucleon-nucleon collision\,\cite{JJ}.

The other chiral-odd distribution $h_{L}$ is twist-3.
Physically higher twist distributions represent complicated
quark-gluon correlations in hadrons.
It is difficult
to isolate them by experiments because they are usually
hidden by the leading twist-2 contributions.
However,
the twist-3 distribution $h_L$
is somewhat immune to this difficulty:
$h_L$ reveals itself as a leading contribution
to the longitudinal-transverse
asymmetry in the Drell-Yan process\,\cite{JJ},
although it appears with
a factor $1/\sqrt{Q^2}$.
This is similar to the
circumstance of
$g_2$ (the transverse spin structure function) whose
contribution to the transversely polarized DIS becomes
leading
order but with the suppression factor
$1/\sqrt{Q^2}$\ \cite{JJ2}.
Thus, the uniqueness of $h_{L}$ is twofold: it is a
``measurable'' higher-twist distribution and corresponds to
chirality violating process. It is expected
to provide new information about the hadron structure and the
QCD dynamics beyond the conventional structure function data.

In addition to more precise measurements of the familiar
distribution functions $f_{1}$ and $g_{1}$,
these ``new'' distribution functions $h_1$, $h_{L}$, and
$g_T$\footnote{$g_{T}$ denotes the twist-3 distribution
corresponding to the structure function $g_{2}$.}
will be measured in the future collider experiments,
such as HERMES, SMC \cite{ad} and RHIC.
In view of this, it is especially important
to develop theoretical study of these distribution functions
as much as possible
based on QCD.
Among these efforts,  the first step is the
perturbative QCD prediction on the
$Q^{2}$-evolution of the distribution functions:
Owing to the factorization property of hard processes,
the $Q^{2}$-evolution of the distribution functions
can be predicted unambiguously in the framework of
the renormalization group and the QCD perturbation theory.
Its prediction is indispensable to extract physical
information
from the experimental data in the high-energy scale
by comparing them with the prediction of hadron models at
the low-energy scale.
Furthermore, the comparison of the $Q^{2}$-evolution itself
between theory and experiment will provide a deeper
test of QCD
beyond the conventional twist-2 level.
The $Q^{2}$-evolution of the twist-2 distribution
functions has
been fully discussed since the first application of QCD
to hard processes;
for example,
the $Q^2$-evolution of $h_1$
was studied in \cite{AM} by employing
the Altarelli-Parisi equation\, \cite{AP}.
As for the higher twist ones,
there has been several works on the
$Q^2$-evolution of $g_T$
by generalizing the Altarelli-Parisi
equation to the higher twist distribution\, \cite{BKL,Rat}
and by
the anomalous dimension calculation\, \cite{JC,ABH,SV,KU}.
But there has been no discussion on the $Q^2$-evolution
of another important twist-3 distribution $h_L$.

In this paper we study the $Q^2$-evolution of $h_L$.
We shall calculate the anomalous dimension matrix for
the twist-3 distribution $h_L$ based on the standard
QCD perturbation theory.
In general, the moments of a twist-3 distribution can be
written
in terms of the matrix elements of
a set of twist-3 operators
involving explicitly the gluon field strength tensor,
and the mixing among them occurs
through renormalization, as was emphasized
in ref. \cite{JC,SV}
in the context of the $g_2$-structure function.
However, the operator mixing occurs not only among these
twist-3
operators
but also with the other twist-3 operators
which vanish by the naive use of the QCD equation of motion,
$(i\rlap/{\mkern-1mu D} - m_q)\psi=0$, (referred to as the
``equation-of-motion (EOM) operators'' from now on).
This is due to the fact that the naive equations of motion
and thus
the vanishing of these operators are not correct as an
operator statement because of quantum effects and
renormalization.
The use of the equations of motion is
allowed only when their matrix elements
are taken with respect to a physical state \cite{P80,JC2}.
On the other hand, the renormalization
of composite operators
has to be carried out in terms of general
Green functions which imbed these
composite operators.
Therefore the mixing involving the EOM operators
is essential to perform renormalization of the
higher twist operators consistently, which was recently
pointed
out by
Kodaira, Yasui, and Uematsu\,\cite{KU} in the context
of $g_2$.
We shall pay particular attention
to this mixing and we will find that it certainly
plays a role
also for the present case of $h_L$.

The outline of this paper is as follows:  In section 2,
we shall
first introduce the chiral-odd distributions
$h_1$ and $h_L$, following the
procedure of ref. \cite{JJ}.
We include this part to
identify the manifestly interaction-dependent operators
(``canonical basis'' \cite{P80,JS})
as well as the EOM operators
relevant for the twist-3 part of $h_L$.
Readers familiar with ref. \cite{JJ}
can skip this part by just noting the existence of the
first term of eq.(\ref{eq210}).
Next we present the general procedure for the renormalization
of the twist-3 operators.
In section 3, we present the actual calculation of
the anomalous dimension
matrix for the twist-3 operators by employing
standard QCD perturbation theory. The calculation
is performed with the Feynman gauge. The loop integration
is dimensionally regularized, and the minimal subtraction
(MS) scheme is adopted.
The details of the calculation
will be discussed in the Appendices to make the
discussion transparent.
In section 4, we will discuss experimental implication of our
result.

\section{Basic formulation}
\setcounter{equation}{0}
\renewcommand{\theequation}{\arabic{section}.\arabic{equation}}

\subsection{Twist-three operators for $h_{L}$}

In this section we shall briefly summarize general
aspects of the
chiral-odd spin-dependent distributions $h_1$ and $h_L$
relevant
for our analysis.  For the detail,
we refer the readers to ref. \cite{JJ}.
The QCD factorization theorem tells us that
a cross section for an inclusive hard process can be
decomposed into the
perturbatively calculable hard cross section and
the parton distribution
function\, \cite{CSS,QS2}.
The latter is known to be written as the light-cone
Fourier transform of the quark (or gluon) correlation
function in a hadron.
The chiral-odd parton distribution functions
(renormalized at the scale $\mu$)
in our interest are defined as follows:
\begin{eqnarray}
\int { d\lambda  \over 2 \pi } e^{i\lambda x} \langle PS |
\bar{\psi}(0) \sigma_{\mu\nu}i\gamma_5
\psi(\lambda n)|_\mu |PS \rangle
= 2 \left[ h_1(x,\mu^2) ( S_{\perp \mu}p_\nu -
S_{\perp \nu}p_\mu )/M
\right.
\nonumber \\
\left.
+ h_L(x, \mu^2)M(p_\mu n_\nu - p_\nu n_\mu )(S \cdot n)
+ h_3(x,\mu^2)M
(S_{\perp \mu} n_\nu -S_{\perp \nu} n_\mu) \right],
\label{eq201}
\end{eqnarray}
where $|PS\rangle$ is the nucleon (mass $M$)
state with its momentum $P$ and spin $S$
($P^{2} = M^{2}$, $S^{2} = -M^{2}$, $P \cdot S=0$).
We introduced the null vectors $p$ and $n$ by the relation
$P_\mu= p_\mu + { M \over 2} n_\mu$, $p^2=n^2=0$, $p\cdot n =1$,
$n^{+} = p^{-} = 0$,
which specify the Lorentz frame of the system.  The
light-cone gauge $n\cdot A =0$ was employed in (\ref{eq201}).
$h_{1}$ and $h_{L}$ are directly accessible by measuring
the proper asymmetries in the polarized Drell-Yan process
\cite{JJ}.
($h_{3}$ is twist-4 and is irrelevant for the following
discussion.)
Taylor expanding the bilocal operator appearing in the
correlation function, l.h.s. of (\ref{eq201}),
one can derive the relation
between
the moment of these parton distribution
functions $h_{1,L}$ and the local operator,
\begin{eqnarray}
\theta^{\mu \nu \mu_1 \cdot\cdot\cdot \mu_n} ={\cal S}_n
\bar{\psi}i\gamma_5 \sigma^{\mu \nu} iD^{\mu_1}
\cdot\cdot\cdot
iD^{\mu_n}
\psi,
\label{eq203}
\end{eqnarray}
where the covariant derivative $D_\mu =
\partial_\mu - igA_\mu$
restores the gauge invariance and
the symbol ${\cal S}_n$ symmetrizes the indices
$\mu_1, ..., \mu_n$.
Here and below, we often suppress the explicit dependence
on the renormalization scale
of the local operators and the parton distribution functions.
For the study of the twist-2
and -3 distributions $h_1$ and $h_L$, it suffices to
consider the piece of
$\theta^{\mu \nu \mu_1 \cdot\cdot\cdot \mu_n}$ which is
further symmetrized among $\nu, \mu_1, ... \mu_n$.
We thus introduce an arbitrary light-like vector $\Delta_\mu$
($\Delta^2=0$) and
consider
\begin{eqnarray}
\theta_n^\mu \cdot \Delta =
\theta^{\mu \nu \mu_1 \cdot\cdot\cdot \mu_n}
\Delta_\nu \Delta_{\mu_1}
\cdot\cdot\cdot \Delta_{\mu_n}.
\label{eq204}
\end{eqnarray}
$\theta_n^\mu \cdot \Delta$ can be decomposed
into the traceless
part $\bar{\theta}$ and the remainder $T$,
\begin{eqnarray}
\theta_n^\mu \cdot \Delta = \bar{\theta}_n^\mu \cdot \Delta
+ T_n^\mu \cdot \Delta
\label{eq205}
\end{eqnarray}
by the condition that
\begin{eqnarray}
g_{\mu\mu_j}\bar{\theta}^{\mu \nu \mu_1 \cdot\cdot\cdot \mu_n}
=g_{\nu\mu_j}\bar{\theta}^{\mu \nu \mu_1
\cdot\cdot\cdot \mu_n}
=g_{\mu_i\mu_j}\bar{\theta}^{\mu \nu \mu_1 \cdot\cdot\cdot
\mu_n}
=0.
\label{eq206}
\end{eqnarray}
$\bar{\theta}^{\mu \nu \mu_1 \cdot\cdot\cdot \mu_n}$
contains all
the twist-2 effect in $\theta^{\mu \nu \mu_1
\cdot\cdot\cdot \mu_n}$
and it is related to the moments of $h_1(x,\mu)$
as
\begin{eqnarray}
{\cal M}_n[h_1(\mu)] &=& a_n(\mu),
\label{eq207}
\\[10pt]
\langle PS | \bar{\theta}_n^\mu \cdot \Delta (\mu)
|PS \rangle &=& {2a_n(\mu) \over M}
\left( S^\mu {\hat P}^{n+1} - P^\mu {\hat S}{\hat P}^n +
{n \over n+2 } M^2 \Delta^\mu {\hat S}{\hat P}^{n-1} \right),
\label{eq208}
\end{eqnarray}
where we introduced the shorthand notation
${\cal M}_n[h(\mu)]\equiv\int dx x^n h(x,\mu)$
and $\hat k \equiv k\cdot \Delta$ for an arbitrary
four vector $k_\mu$.
Also, we obtain for the moments of $h_{L}$:
\begin{eqnarray}
{\cal M}_{n}[h_{L}] = \frac{2}{n+2} {\cal M}_{n}[h_{1}]
+ {\cal M}_{n}[\widetilde{h}_{L}].
\label{eqnew1}
\end{eqnarray}
The first term shows that $h_{L}$ receives a contribution
from the twist-2 distribution $h_{1}$. This piece
is an analogue
of the Wandzura-Wilczek contribution \cite{WW} for $g_{2}$.
${\cal M}_{n}[\widetilde{h}_{L}]$ of (\ref{eqnew1})
are directly
related to the matrix elements of the
twist-3 operator $T_{n}^{\mu} \cdot \Delta$
of (\ref{eq205}).
Explicit calculation gives
\begin{eqnarray}
T_n^\mu \cdot \Delta = { \Delta^\mu \over n+2 }
\sum_{j=0}^{n-1} \bar{\psi} i \gamma_5 \sigma^{\rho \nu}
\Delta_\nu d^j iD_\rho d^{n-j-1} \psi,
\label{eq209}
\end{eqnarray}
where $d \equiv iD\cdot \Delta$.
$T_n^\mu \cdot \Delta$ can be recast
into the following form
by using the relation $[D_\mu, D_\nu] = -igG_{\mu\nu}$
with $G_{\mu\nu}$ the gluon field strength tensor
\begin{equation}
T_n^\mu \cdot \Delta = {n \over n+2} \Delta^\mu E_n
\cdot \Delta
+ {n \over n+2 } \Delta^\mu N_{n}\cdot\Delta
- \Delta^\mu \sum_{l=2}^{ \left[ {n+1 \over 2}
\right] } \left( 1 - {2l \over n+2 } \right) R_{n,l}
\cdot \Delta,
\label{eq210}
\end{equation}
where ${\cal O}\cdot\Delta \equiv
{\cal O}^{\mu_{1}\cdots\mu_{n}}\Delta_{\mu_{1}}
\cdots \Delta_{\mu_{n}}$.
Here the first term (the ``EOM operator'')
is defined as
\begin{equation}
E_n^{\mu_1
\cdot\cdot\cdot \mu_n}
= {1 \over 2} {\cal S}_{n}
\left[ \bar{\psi}(i\rlap/{\mkern-1mu D} -m_q )
\gamma_5 \gamma^{\mu_{1}}
iD^{\mu_{2}} \cdots iD^{\mu_{n}} \psi +
\bar{\psi} \gamma_5 \gamma^{\mu_{1}}
iD^{\mu_{2}} \cdots iD^{\mu_{n}}(i\rlap/{\mkern-1mu D} -m_q
)
\psi \right]
- {\rm traces}.
\label{eq211}
\end{equation}
This operator vanishes by the naive use of
the QCD equation of motion $(i\rlap/{\mkern-1mu D}
-m_q)\psi=0$.
We can set it to zero when we take its matrix element
with respect to a physical state (such as the nucleon state)
\cite{P80,JC2}, which is why it is discarded in ref.\cite{JJ}.
However, this is not an operator identity and the
mixing
between $E_{n}$ and the other twist-3 operators defined below
should be taken into account during the course of
renormalization.
The second term is given by
\begin{equation}
N_{n}^{\mu_{1} \cdots \mu_{n}}
= {\cal S}_{n} m_q \bar{\psi} \gamma_5 \gamma^{\mu_{1}}
iD^{\mu_{2}} \cdots iD^{\mu_{n}} \psi - {\rm traces}.
\label{eq211n}
\end{equation}
$R_{n,l}^{\mu_1 \cdot\cdot\cdot \mu_n} $ in the third term
of
(\ref{eq210})
is defined as
\begin{eqnarray}
R_{n,l}^{\mu_1 \cdot\cdot\cdot \mu_n}
&=& \theta_{n-l+2}^{\mu_1 \cdot\cdot\cdot \mu_n}
- \theta_l^{\mu_1 \cdot\cdot\cdot \mu_n}, \ \ \ \left(
l=2,...,\left[{n+1 \over 2} \right] \right)
\label{eq212} \\[10pt]
\theta_{l}^{\mu_1 \cdot\cdot\cdot \mu_n}
&=& {1 \over 2}
{\cal S}_n \bar{\psi} \sigma^{\alpha \mu_1} i \gamma_5
iD^{\mu_2} \cdot\cdot\cdot igG^{\mu_l}_{\ \ \alpha}
\cdot\cdot\cdot
iD^{\mu_n} \psi - {\rm traces},
\label{eq213}
\end{eqnarray}
which explicitly involves the gluon field strength tensor,
suggesting that the twist-3 operators truly represents
the effect of quark-gluon correlations.
By the combination of $\theta_{l}^{\mu_1 \cdot\cdot\cdot
\mu_n}$
in the form of the r.h.s. of (\ref{eq212}),
$R_{n,l}^{\mu_1 \cdot\cdot\cdot \mu_n}$
can have definite charge conjugation property.

With these definitions and the relation (\ref{eq210}),
the $n$-th moment of the genuine twist-3 piece of
$h_L$ can be written down
in terms of the
nucleon matrix elements of the twist-3 operators
(see (\ref{eqnew1}))
\begin{eqnarray}
{\cal M}_n [\widetilde{h}_L] =
{n \over n+2}{m_q \over M}{\cal M}_{n-1}[g_1]
+ {\cal M}_n[h_L^3].
\label{eq214}
\end{eqnarray}
The first term is due to the contribution
of the second term of eq.(\ref{eq210}),
and thus shows the quark mass effect;
it is in fact the quark mass
times the twist-2 operator corresponding
to the $g_{1}$-distribution:
\begin{eqnarray}
\langle PS |{\cal S}_{n}\bar{\psi}\gamma^{\mu_1}\gamma_5
iD^{\mu_2}\cdot\cdot\cdot iD^{\mu_n}\psi |PS \rangle
=2 {\cal M}_{n-1}[g_1(\mu)]{\cal S}_{n}(S^{\mu_1}P^{\mu_2}
\cdot\cdot\cdot
P^{\mu_{n}} - {\rm traces}).
\label{eq215}
\end{eqnarray}
The second term designates the contribution from
$R_{n,l}^{\mu_1 \cdot\cdot\cdot \mu_n} $:
\begin{eqnarray}
{\cal M}_{n}[h_L^3] =
\sum_{l=2}^{ \left[ {n+1 \over 2}
\right] } \left( 1 - {2l \over n+2 } \right) b_{n,l}(\mu^2)
\label{eq216}
\end{eqnarray}
with
\begin{eqnarray}
\langle PS | R_{n,l}^{\mu_1\cdot\cdot\cdot \mu_n}(\mu^2)
|PS \rangle
= 2 b_{n,l}(\mu^2) M {\cal S}_n(S^{\mu_1}P^{\mu_2}
\cdot\cdot\cdot P^{\mu_n}
-{\rm traces}).
\label{eq217}
\end{eqnarray}
By inverting the moments,
(\ref{eqnew1}) and (\ref{eq214}) give the relation between the
structure functions
themselves:
\begin{eqnarray}
h_L(x, \mu^2) = 2x \int_x^1 { h_1(y,\mu^2) \over y^2 } dy
+ {m_q \over M}\left[ {g_1(x,\mu^2) \over x } - 2x \int_x^1
{ g_1(y,\mu^2) \over y^3 } dy \right] + h_L^3(x,\mu^2),
\label{eq218}
\end{eqnarray}
for $x>0$.\footnote{$h_{L}(x)$ for $x <0$
should be related to the ``antiquark distribution''
$\overline{h}_{L}(x)$ as $\overline{h}_{L}(x) = - h_{L}(-x)$
by charge conjugation.}

\subsection{Renormalization of the twist-three operators}

We now proceed to discuss the renormalization of the twist-3
operators.
As discussed in section 2.1,  $[(n+1)/2]+1$ twist-3 operators,
$R_{n,l}\cdot\Delta$ ($l=2,3,...\left[{n+1 \over 2}\right]$)
$N_n\cdot \Delta$ and
$E_n\cdot\Delta$ formally participate in the
$n$-th moment of ${\tilde h}_L(x,Q^2)$.
It has been known \cite{JS} that as a complete basis of
higher twist operators
one can always choose ``canonical" ones which (1) are
traceless and symmetric with respect to all the
Lorentz indices and (2) have no contracted derivatives.
Any noncanonical operators
which could appear through radiative corrections
can be transformed into the canonical
ones modulo EOM operators
by use of the relation $\left[ D_\mu, D_\nu \right] = -ig
G_{\mu\nu}$,
and the physical matrix elements of the EOM operators vanish.
These canonical operators mix with each other and also with
the EOM operators under renormalization.
Therefore we can choose
$R_{n,l}\cdot\Delta$, $N_n\cdot \Delta$,
and $E_{n} \cdot \Delta$
as a basis for renormalization.
We further recall that the renormalization
of composite operators generally involves the mixing with
gauge noninvariant EOM operators
which do not exist in the original basis\,\cite{JC2}.
We will come back to this point in the next section.

The scale dependence of the physical matrix elements,
e.g., $b_{n,l}(\mu)$ of (\ref{eq217})
are determined by the anomalous dimensions of the
corresponding
composite operators.
To see this, we write down the renormalization group equation
for these operators.
The bare- (${\cal O}_{i}^{B}$)
and the renormalized- (${\cal O}_{i}$)
composite operators are related by the renormalization
constant matrix
$Z_{ij}$:
\begin{eqnarray}
{\cal O}_i(\mu)=Z^{-1}_{ij}(\mu) {\cal O}^B_j,
\label{eq219}
\end{eqnarray}
where ${\cal O}_i$ symbolically refer to
$R_{n,l}^{\mu_1,\cdot\cdot\cdot\mu_n}$,
$E_n^{\mu_1,\cdot\cdot\cdot\mu_n}$, and $N_{n}^{\mu_{1},
\cdot\cdot\cdot\mu_{n}}$.
The renormalization group equation for ${\cal O}_{i}(\mu)$
is obtained
by using the fact that the unrenormalized operators
do not depend on the
renormalization scale:
\begin{eqnarray}
\mu {d{\cal O}_{i}(\mu) \over d \mu }
+ \gamma_{ij}\left(g(\mu)\right)
{\cal O}_{j}(\mu)
=0,
\label{eq221}
\end{eqnarray}
where the anomalous dimension matrix $\gamma_{ij}$
for $\{R, E, N\}$
is defined as
\begin{eqnarray}
\gamma_{ij}(\mu)=\mu {d Z_{kj}(\mu) \over d\mu }
Z^{-1}_{ik}(\mu).
\label{eq222}
\end{eqnarray}
In the leading logarithmic approximation, this equation
is solved to give
\begin{eqnarray}
{\cal O}_{i}(Q^2) = \sum_{j} \left[
\left( { \alpha(Q^2) \over \alpha(\mu^2)}
\right)^{ \gamma_0 \over 2\beta_0 }
\right]_{ij} {\cal O}_{j}(\mu^2),
\label{eq223}
\end{eqnarray}
where $\alpha(\mu^2)$ is the QCD running coupling constant,
$\beta_0=\left(11-(2/3)N_f\right) /(4 \pi)^{2}$
and $\gamma_{0}$
are the lowest order (1-loop) coefficient of
the $\beta$-function and
the anomalous dimension matrix
\begin{eqnarray}
\gamma_{ij}\left(g(\mu)\right) = \gamma_{ij}^0
g(\mu)^2 + O(g(\mu)^4).
\label{eq224}
\end{eqnarray}

In order to compute the $Z_{ij}$-factor
of (\ref{eq219}),
one imbeds the composite operators ${\cal O}_{i}$ into an
appropriate Green functions with a convenient kinematics,
and computes radiative corrections to this Green function.
For a familiar case of the twist-2 operators, for example,
the two-point functions with the on-shell external lines
are usually
considered.
In the present case, however,
the EOM operator $E_{n}^{\mu_{1},\cdot\cdot\cdot\mu_{n}}$
should be retained as a nonzero quantity
and the mixing of them with the other operators
should be consistently taken into account:
This means that
it is necessary to compute Green functions with
the off-shell kinematics for the external lines.

Thus, in order to compute renormalization of
${\cal O}_{i} = R_{n,l}^{\mu_1,\cdot\cdot\cdot\mu_n}$,
for example,
we may consider the truncated
three-point Green function $F_{i}(p,q,q-p)$ defined by
\begin{eqnarray}
& &  F_{i}(p,q,k)(2\pi)^4\delta^4(p+k-q)
G(p)G(q)D(k) \nonumber\\[10pt]
& & \ \ \ \ =
\int d^4x\, d^4y\, d^4z\, e^{ipx} e^{-iqy} e^{ikz}
\langle T \{ {\cal O}_i \psi(x) \bar{\psi}(y) A_\mu(z) \}
\rangle, \ \ \ \ \
\label{eq226}
\end{eqnarray}
where $G$ and $D$ are the quark and gluon propagators,
respectively. (We suppressed the Lorentz and
the spinor indices for simplicity.)
We consider the three-point Green function for the
off-shell quark and gluon external lines
(not a physical state!), and therefore $F_i$ with
${\cal O}_i = E_n$
do not vanish. In the next section,
we present the one-loop calculation of
this function to get $Z_{ij}$.

\section{Anomalous Dimension Matrix for Twist-3 Operators}
\setcounter{equation}{0}
\label{sec3n}
\renewcommand{\theequation}{\arabic{section}.\arabic{equation}}

In this section we present the computation of the
anomalous dimension matrix for the twist-3 operators
for $h_{L}$.
The calculation is performed up to the one-loop order.
The loop integrals are dimensionally regularized and
the MS scheme is employed for renormalization.
Thus we keep only the simple dimensional pole
proportional to $1 /\varepsilon$ in the one-loop amplitudes
($\varepsilon =  (4-d) / 2$ with $d$ the space-time
dimension).
We use the Feynman gauge for the gluon propagator,
but the results should be independent
of the gauge.

As we discussed in the last section,
we imbed the relevant twist-three composite operators
into the three-point function $F_{i}(p, q, k)$ of (\ref{eq226})
assuming the off-shell kinematics.
In this case, the basic ingredients are the tree level
vertices
for the operators $R_{n,l}$, $E_{n}$, $N_{n}$
corresponding to the diagrams shown in Fig. 1,
which we call the ``basic vertices''.

The three-point basic vertex of $R_{n,l}\cdot\Delta$
shown in Fig. 1 (a) becomes
\begin{eqnarray}
{\cal R}^{(3)}_{n, l, \mu} =
{g \over 2} \sigma^{\alpha\lambda}\Delta_\lambda i\gamma_5
({\hat p}^{n-l}{\hat q}^{l-2}-{\hat p}^{l-2}{\hat q}^{n-l})
( -{\hat k} g_{\alpha\mu}
+ k_\alpha \Delta_\mu ) t^a,
\label{eq301}
\end{eqnarray}
while those for $E_{n}\cdot\Delta$, $N_{n}\cdot \Delta$
are given by
\begin{eqnarray}
{\cal E}^{(3)}_{n, \mu}
& =&{g \over 2} \left[ \gamma_\mu \gamma_5 \rlap/{\mkern-1mu
\Delta}
{\hat q}^{n-1}
+\gamma_5 \rlap/{\mkern-1mu \Delta} {\hat p}^{n-1}
\gamma_\mu
+ \Delta_\mu \sum_{j=2}^n
(\rlap/{\mkern-1mu p} -m_q) \gamma_5 \rlap/{\mkern-1mu
\Delta}
{\hat p}^{j-2} {\hat q}^{n-j}
\right.\nonumber\\
& & \left.
\ \ \ \ \ \ + \Delta_\mu \sum_{j=2}^n\gamma_5
\rlap/{\mkern-1mu \Delta} (\rlap/{\mkern-1mu q} -m_q)
{\hat p}^{j-2}  {\hat q}^{n-j}
\right]t^a ,
\label{eq303}\\
{\cal N}^{(3)}_{n, \mu}
& =& m_q g\Delta_{\mu} \gamma_{5} \rlap/{\mkern-1mu \Delta}
\sum_{j = 2}^{n}
{\hat p}^{j-2}{\hat q}^{n-j} t^{a},
\label{eq303n}
\end{eqnarray}
where $k=q-p$ and
$t^a$ ($a=1,...,{N_c^2 -1}$) is the color matrix
normalized as ${\rm Tr}(t^at^b) = {1 \over 2}\delta^{ab}$.

We first imbed $R_{n,l}\cdot\Delta$ into the three-point
function $F_{i}$ of (\ref{eq226}). The Feynman diagrams
which give the one-loop radiative corrections to the operator
$R_{n,l}\cdot\Delta$ are the one-particle-irreducible diagrams
shown in Fig. 2. In order to write down
those amplitudes, the vertex for $R_{n,l}\cdot\Delta$
corresponding
to Fig. 1(b) is necessary in addition to the usual Feynman rules:
\begin{eqnarray}
& & {g^2 \over 2} \left[ if^{abc}t^c \sigma^{\alpha\lambda}
\Delta_\lambda i\gamma_5
\Delta_\mu g_{\alpha\nu} {\hat p}^{n-l} {\hat q}^{l-2}
\right. \nonumber\\[8pt]
& & \left.
+ \sum_{j=2}^{n-l+1}
\sigma^{\alpha\lambda}
\Delta_\lambda i\gamma_5
t^at^b{\hat p}^{j-2} \Delta_\mu ({\hat p} + {\hat k})^{n-l+1-j}
(-{\hat k'}g_{\alpha\nu} + k'_\alpha \Delta_\nu ) {\hat q}^{l-2}
\right. \nonumber\\[8pt]
& & \left.
+ \sum_{j=n-l+3}^n
\sigma^{\alpha\lambda}
\Delta_\lambda i\gamma_5
t^at^b{\hat p}^{n-l} \Delta_\nu ({\hat p} + {\hat k})^{j-n+l-3}
 (-{\hat k}g_{\alpha\mu} + k_\alpha \Delta_\mu ) {\hat q}^{n-j}
\right. \nonumber\\[8pt]
& & \left. + (\mu \leftrightarrow \nu, k \leftrightarrow k',
a \leftrightarrow b)
\right]  - ( l \rightarrow n-l+2)
\label{eq302}
\end{eqnarray}
with $k'=q-p-k$.
We have to add the counter term contribution
(see (\ref{eq219})),
\begin{equation}
\left(Z^{-1}_{ll'}Z_{2}\sqrt{Z_{3}}Z_{g} - \delta_{ll'} \right)
{\cal R}^{(3)}_{n, l', \mu}
+Z^{-1}_{lE}Z_{2}\sqrt{Z_{3}}Z_{g}
\left\{{\cal E}^{(3)}_{n, \mu}
+\left(Z_{m}-1\right){\cal N}_{n,\mu}^{(3)}\right\}
+ Z^{-1}_{lN}Z_{2}\sqrt{Z_{3}}Z_{g}Z_{m} {\cal N}^{(3)}_{n, \mu}
\label{eq302n}
\end{equation}
to the sum of all the one-loop amplitudes of Fig. 2,
and require that the total results be finite as $\varepsilon
\rightarrow 0$. In (\ref{eq302n}), $Z_{2}$ and $Z_{3}$
are the usual wave function renormalization constants
for the quark and gluon fields,
while $Z_{g}$ and $Z_{m}$ are the coupling constant and mass
renormalization constants defined by
\begin{eqnarray}
g = \frac{1}{Z_{g}} \mu^{- \varepsilon}g^{B} ;
\;\;\;\;
m_q = \frac{1}{Z_{m}} m^{B}_q,
\label{eq302nn}
\end{eqnarray}
where ``B'' denotes the unrenormalized quantities
similarly to (\ref{eq219}).
Note that ${\cal R}^{(3)}_{n, l, \mu}$,
${\cal E}^{(3)}_{n, \mu}$
and ${\cal N}_{n,\mu}^{(3)}$
are proportional to the coupling constant and thus each term
of (\ref{eq302n}) involves the factor $Z_{g}$
as a coefficient.
Similarly, the factor $Z_{m}$ appears because
${\cal N}^{(3)}_{n, \mu}$ contains the quark mass $m$.
In the MS scheme, (the finite part of) the
renormalization constants are chosen so that the counter term
contributions (\ref{eq302n}) precisely cancel out the terms
proportional to the $1/\varepsilon$ pole
from the dimensionally regularized one-loop amplitudes.

For the case of the three-point functions
imbedding $E_{n}\cdot\Delta$ or $N_{n}\cdot\Delta$,
we can proceed in a similar manner
by interchanging the roles of $R_{n,l}\cdot\Delta$
with $E_{n}\cdot\Delta$
or $N_{n}\cdot\Delta$.
These results, together with the well known results
for $Z_{2,3,g, m}$ (in the Feynman gauge),
\begin{eqnarray}
Z_2 = 1-{ g^2 \over (4\pi)^2 \varepsilon } C_F; \;\;
Z_g \sqrt{Z_3} = 1-{ g^2 \over (4\pi)^2 \varepsilon } C_G; \;\;
Z_{m} = 1 - 3\frac{g^{2}}{(4\pi)^{2} \varepsilon}C_{F}
\label{eq308n}
\end{eqnarray}
completely determine the relevant $Z_{ij}$-factor
($i, j=2,...,\left[{n+1 \over 2}\right], E, N$).
($C_F = {N_c^2 -1 \over 2N_c}$ and $C_G = N_c$ are the Casimir
operators of the color gauge group SU($N_{c}$).)

The actual computation of the Feynman amplitudes of Fig. 2
is rather cumbersome due to the complicated structure
of the vertices,
e.g., (\ref{eq301})-(\ref{eq302}).
Moreover, the participation of the gauge-noninvariant
EOM operators
would further complicate the computation
(see below).
In this light
it is convenient to employ the following procedure (1)-(3)
to make the computation tractable:

(1) We introduce a vector $\Omega_{\mu}$,
which satisfies the condition
$\hat{\Omega}= \Delta^{\mu} \Omega_{\mu}= 0$.
The contraction of the implicit Lorentz index $\mu$ of $F_{i}$
((\ref{eq226})) with $\Omega_{\mu}$ kills off many terms
and simplifies
the computation enormously: For example, the basic vertices in
(\ref{eq301}) and (\ref{eq303}) become
\begin{eqnarray}
{\cal R}_{n,l}^{(3)}\cdot \Omega
= -{g \over 2} \Omega_\alpha \sigma^{\alpha\lambda}
\Delta_\lambda i \gamma_5 ({\hat q}-{\hat p}) \left(
{\hat p}^{n-l} {\hat q}^{l-2} - {\hat p}^{l-2}{\hat q}^{n-l}
\right)t^a
\label{eq304}
\end{eqnarray}
and
\begin{eqnarray}
{\cal E}_n^{(3)}\cdot\Omega
={g \over 2} \Omega_\alpha \sigma^{\alpha\lambda}
\Delta_\lambda i \gamma_5 \left(
{\hat p}^{n-1} + {\hat q}^{n-1} \right) t^a.
\label{eq305}
\end{eqnarray}
This contraction brings another favorable effect:
As was exemplified in ref.\cite{KU}
in the context of $g_T$, the gauge-noninvariant EOM operators
generally mix through renormalization
in addition to the gauge-invariant EOM operator
$E_{n}$\,\cite{JC2}.
Typically, those gauge-noninvariant operators are obtained
by replacing
some of the uncontracted covariant derivatives
contained in $E_{n}^{\mu_{1} \cdots \mu_{n}}$
by the simple derivatives.
Therefore, for large $n$,
a large number of different gauge-noninvariant operators are
expected to come
into play.
Due to this phenomenon, it is extremely difficult
to identify the tensor structure
of the one-loop amplitudes
by the basic vertices (\ref{eq301})-(\ref{eq303n})
and those for the
gauge-noninvariant EOM operators.
On the other hand,
after the contraction with $\Omega_{\mu}$,
we do not distinguish $E_{n}\cdot\Delta$
and the gauge-noninvariant EOM
operators; i.e.,
the contracted basic vertex ${\cal E}^{(3)}_{n}\cdot\Omega$
of (\ref{eq305}) and the contracted basic vertices for the
gauge-noninvariant EOM operators coincide due to the condition
$\hat{\Omega} = 0$. It is easy to see that
such an identification
between the gauge-invariant and gauge-noninvariant EOM operators
does not affect the prediction of the $Q^{2}$-evolution of the
moment sum rules, because of the property that the physical
matrix elements of the EOM operators vanish.

(2) For the computation of the three-point functions imbedding
$R_{n,l}\cdot\Delta$ and $E_{n}\cdot\Delta$, we set $m_{q} = 0$.
Taking this limit is legitimate for the present case
employing the off-shell kinematics for the external lines,
and amounts to neglecting the contribution of the basic
vertex ${\cal N}_{n, \mu}^{(3)}$ of (\ref{eq303n})
in the preceding discussions of this section.
Clearly, this procedure still gives the correct results for the
renormalization mixing between $R_{n,l}$ and $E_{n}$, i.e.,
for $Z_{ij}$ with $i,j = 2, \cdots , [(n+1)/2], E$.
Actually, we need not compute the one-loop correction to the
three-point functions imbedding $E_{n}\cdot\Delta$:
The property $\langle PS | E_n \cdot \Delta | PS \rangle =0$
immediately implies
$Z_{Ei}= 0$ ($i = 2, \cdots, [(n+1)/2], N$).

(3) In order to obtain the other components
of $Z_{ij}$, it is sufficient to consider the
two-point functions
shown in Fig. 3 with the insertion of the relevant operators.
We again employ the off-shell kinematics
for the external quark lines,
but the computation is now performed for the
nonzero quark mass.
The basic vertices corresponding to the operators
$E_{n}\cdot\Delta$,
$N_{n}\cdot\Delta$ are given by
\begin{eqnarray}
{\cal E}_n^{(2)}&=& {1 \over 2} {\hat p}^{n-1}
\left( \rlap/{\mkern-1mu \Delta} \rlap/{\mkern-1mu p}
- \rlap/{\mkern-1mu p} \rlap/{\mkern-1mu \Delta}
+ 2m_q \rlap/{\mkern-1mu \Delta} \right)
\gamma_5,
\label{eq309}\\
{\cal N}_n^{(2)}&=& m_q \gamma_5\rlap/{\mkern-1mu \Delta}
{\hat p}^{n-1}
\label{eq310},
\end{eqnarray}
while the one for $R_{n,l}\cdot\Delta$ vanishes.
Thus, the computation of the one-loop corrections
to the two-point
functions imbedding $R_{n,l}$, $E_{n}$
and $N_{n}$,
by using the vertex given in
(\ref{eq301})-(\ref{eq303n})
and by following the steps
similar to the case of the three-point functions,
gives the components
$Z_{iE}$, $Z_{iN}$ ($i = 2, \cdots, [(n+1)/2], E, N$).
For the remaining components $Z_{Ni}$ ($i = 2, \cdots,
[(n+1)/2]$)
we need not perform any actual computation:
Those vanish because $N_{n}$ is a twist-2 operator
multiplied by a
quark mass.

We note that the several components of $Z_{ij}$
can be obtained
by different methods,
giving a consistency check of our procedure:
$Z_{iE}$ ($i = 2, \cdots, [(n+1)/2]$) by calculating the two-
as well as three-point functions;
$Z_{EN}$, $Z_{NE}$ by the calculation of the two-point
functions
and by a special property of the operators $E_{n}$, $N_{n}$
discussed above.
For all those cases, the different methods gave the
identical results.

In the above discussion, we completely neglected the
flavor structure
of the operator.  Even for the case of flavor-singlet
combinations
of $h_{1,L}$, there exist no gluon distributions
which mix with
them.   In fact, if there would be any mixing between
flavor-singlet
$h_1$ and a gluon distribution, it would arise from the
diagrams
shown in Fig.4.
But all of them are identically zero because of chirality.
The situation for $h_L$ is completely the same.

Now we get all the $Z_{ij}$-factor
for the twist-3 operators.
(The contributions of the relevant three- and two-point
Feynman amplitudes corresponding to the diagrams of Figs.2
and 3 are
presented in Appendix A.)
We summarize the final result in the following
matrix form:
\begin{eqnarray}
\left(\matrix{R_{n,l}^B\cr
              E_n^B\cr
              N_n^B\cr}\right)=
\left(\matrix{Z_{lm}(\mu)&Z_{lE}(\mu)&Z_{lN}(\mu)\cr
              0&Z_{EE}(\mu)&0\cr
              0&0&Z_{NN}(\mu)\cr}\right)
\left(\matrix{R_{n,m}(\mu)\cr
              E_n(\mu)\cr
              N_n(\mu)\cr}\right),
\ \ \ \left(l,m = 2,\cdot\cdot\cdot,\left[{n+1 \over 2}
\right]\right).
\label{eq311}
\end{eqnarray}
If we express $Z_{ij}$ as
\begin{eqnarray}
Z_{ij} = \delta_{ij}+ {g^2 \over 16\pi^2 \varepsilon} X_{ij}
\ \ \ \ \left(i,j= 2,\cdot\cdot\cdot,\left[{n+1 \over 2}
\right],
E, N\right),
\label{eq312}
\end{eqnarray}
then $X_{ij}$ is given as follows:
\begin{eqnarray}
X_{lm} &=&  C_G
\left[
 {m+1 \over 2}
\left(
{1 \over [n-l+1]_2 } -
{ 1 \over [l-1]_2} \right)
 \right. \nonumber\\[8pt]
& & \left.
+ 2 \left(
{ 1 \over n-l+2} - {1 \over l}
\right)
  - {1 \over n-l+2-m } + { 1 \over l-m}
\right. \nonumber\\[8pt]
& & \left.
+ {1 \over 2} \left(
{2l-1 \over [l-1]_2 } - {m(l-3) \over [l-1]_3 }
\right)
- {1 \over 2}
\left(
{2n-2l+3 \over [n-l+1]_2}
-{m(n-l-1) \over [n-l+1]_3 }
\right)
\right]  \nonumber\\[8pt]
& &+ (C_G -2C_F) \left[
{ (-1)^{n-l-m}\ _{n-l}C_{m-2} \over
(n-l+2-m)\ _{n-m+1}C_{l-1}  }
- { (-1)^{l-m}\ _{l-2}C_{l-m} \over
(l-m)\ _{n-m+1}C_{l-m} }
\right. \nonumber\\[8pt]
& & \left. + 2 (-1)^m \left( { _{l-1}C_{m-1} \over [l-1]_3 }
- { _{n-l+1}C_{m-1} \over [n-l+1]_3 } \right)  \right]
 \ \ \ \ \ (2 \le m \le l-1),
\label{eq313}
\end{eqnarray}
\begin{eqnarray}
X_{ll} &=& C_G \left[ - { (l-1)(n+2) \over 2(n-2l+2)
[n-l+1]_2 }
-{1 \over 2} \left( S_{n-l+1} + S_{l-1} + S_{n-l+2}
+ S_l \right)
\right.
\nonumber\\[8pt]
& &\left.
+{1 \over 2} \left( { (2l-n-2)(n-l-1) \over [n-l+1]_3 }
+ { l+2 \over l(l-1) } + {l-3 \over [l-1]_3 } \right)
- \frac{1}{l}
\right]
\nonumber \\[8pt]
& &+2(C_G-2C_F)\left[(-1)^l \left( {1 \over [l-1]_3 }+
{ (-1)^n -\ _{n-l+1}C_{l-1} \over [n-l+1]_3 } \right)\right.
\nonumber \\[8pt]
& &\left.+(-1)^{n} { l-1 \over 2(n-l+1)(n-2l+2) }\right]
\nonumber \\[8pt]
& &-C_F \left( 2S_{l-1} + 2S_{n-l+1} -3 \right),
\label{eq314}
\end{eqnarray}
\begin{eqnarray}
X_{lm} &=& C_G \left[ - { 1 \over n-l+2 -m} - {1 \over l-m}
-{ n-2m+2 \over 2[n-l+1]_2 }
\right. \nonumber\\[8pt]
& & \left. + { (n-l-1)(2m-n-2) \over 2[n-l+1]_3 }
\right] \nonumber\\[8pt]
& & - (C_G-2C_F) \left[{ 2 (-1)^m \left( _{n-l+1}C_{m-1}
- (-1)^n
\ _{n-l+1}C_{m-l}
\right) \over
[n-l+1]_3 }\right.
\nonumber\\[8pt]
& & \left.
+{ (-1)^{m-l}\ _{n-l}C_{m-l} \over (m-l)\ _{m-1}C_{m-l} }
- { (-1)^{n-l-m}\ _{n-l}C_{m-2}
\over (n-l+2-m)\ _{n-m+1}C_{l-1} }
\right]
\nonumber\\[8pt]
& & \ \ \ \ \ \ \ \ \ \ \ \
\left( l+1 \le m \le \left[ {n+1 \over 2} \right] \right),
\label{eq315}
\end{eqnarray}
\begin{eqnarray}
X_{lE} =  2C_F \left( {1 \over [l]_{2} } - { 1 \over [n-l+2]_2}
 \right),
\label{eq316}
\end{eqnarray}
\begin{eqnarray}
X_{lN}= -4C_F \left( {1 \over [l-1]_3 } - {1 \over [n-l+1]_3}
\right),
\label{eq317}
\end{eqnarray}
\begin{eqnarray}
X_{EE} = 2 (1-S_n)C_F,
\label{eq318}
\end{eqnarray}
\begin{eqnarray}
X_{NN}=C_F\left({2 \over n(n+1) } -4S_n \right),
\label{eq320}
\end{eqnarray}
where $S_{n} = \sum_{j=1}^{n}1/j$
and $[j]_{k} = j(j+1)\cdots(j+k-1)$.
With these $X_{ij}$
($i,j = 2,3,...,\left[ {n+1 \over 2} \right], E, N$),
the anomalous dimension matrix for the twist-3 operators
$R_{n,l}$, $E_n$ and $N_n$ take the form of
the upper triangular matrix as
\begin{eqnarray}
\gamma_{ij} = - {g^2 \over 8\pi^2 } X_{ij}.
\label{eq321}
\end{eqnarray}
 From (\ref{eq223}),
the $Q^2$-evolution of the nucleon matrix elements
are given by
\begin{eqnarray}
b_{n,l}(Q^2) &=&
\sum^{[\frac{n+1}{2}]}_{m=2}\left[ \left(
{\alpha(Q^2) \over \alpha(\mu^2)}
\right)^{-X /16\pi^2\beta_0} \right]_{lm}b_{n,m}(\mu^2)
+ \left[ \left(
{\alpha(Q^2) \over \alpha(\mu^2)}
\right)^{-X /16\pi^2\beta_0} \right]_{lN}d_{n}(\mu^2),
\label{eq322}\\
d_{n}(Q^{2}) &=& \left(
{\alpha(Q^2) \over \alpha(\mu^2)}
\right)^{-X_{NN} /16\pi^2\beta_0} d_{n}(\mu^2),
\label{eq322n}
\end{eqnarray}
where we set $d_{n}(\mu^{2}) = (m_q/M){\cal M}_{n-1}
[g_{1}(\mu^{2})]$
(see (\ref{eq215})).
This is our main result of this paper.

Before leaving this section, we shall briefly discuss
the $Q^2$-evolution
of the twist-2 distribution $h_1(x,Q^2)$ because it appears
in $h_{L}$ as a Wandzura-Wilczek analogue ((\ref{eqnew1}) and
(\ref{eq218})).
There is
only one operator
$\bar{\theta}_{n}^{\mu}\cdot \Delta$ (eq.(\ref{eq205}))
for each $n$ in the twist-2 level,
and thus there is no complication arising from operator mixing.
The lowest order coefficient of the anomalous dimension
can be obtained from the one-loop diagram of the two-point
functions
shown in Fig. 3.
We note that
the contribution from the diagram in Fig.3 (a) vanishes,
and the contribution to the $Z$-factor
from Figs. 3 (b) and (c) is the same as those for $f_1$
and $g_1$
from the same diagrams.
By using a similar technique as above,
we obtain for the $Z$-factor for the composite operator
$\bar{\theta}_{n}^{\mu}\cdot \Delta$,
which is now a single constant and not a matrix:
\begin{eqnarray}
Z = 1 - { g^2 \over 16 \pi^2 \varepsilon }C_F K_{n+1},
\label{eq228}
\end{eqnarray}
with
\begin{eqnarray}
K_n= 1 + 4 \sum_{j=2}^{n} {1 \over j}.\nonumber\\
\label{eq229}
\end{eqnarray}
By substituting the result into (\ref{eq222}),
we get the anomalous dimension
\begin{eqnarray}
\gamma = {g^2 \over 8\pi^2}C_FK_{n+1}.
\label{eq230}
\end{eqnarray}
This governs the $Q^{2}$-evolution of the first term
on the r.h.s.
of (\ref{eqnew1})
(see eqs. (\ref{eq207}) and (\ref{eq208})).
The same result was obtained in ref.\cite{AM} from the
Altarelli-Parisi equation for $h_1$.

\section{Examples of $Q^2$-Evolution}
\setcounter{equation}{0}
\renewcommand{\theequation}{\arabic{section}.\arabic{equation}}
Here we present
some examples of the $Q^2$-evolution of
the twist-3 distribution ${\tilde h}_L(x,Q^2)$
by using the result
obtained in the previous section.  Since there is no mixing
of $h_L$ with the gluonic distribution,
we shall consider a distribution for one
quark flavor.  We also neglect the
contribution from the quark mass operator $N_{n}$ in the
following.

For the 3-rd and 4-th moments of ${\tilde h}_L$,
only one twist-3
operator $R_{n,l}$
contributes.  Their anomalous dimensions are
(ignoring the common factor
$g^2/8\pi^2$) 104/9 and 1099/90 for the 3-rd and 4-th
moments,
respectively.
This gives for ${\cal M}_{3,4}[{\tilde h}_L(Q^2)]$ as
\begin{eqnarray} {\cal M}_3 [{\tilde h}_L(Q^2)] ={1
\over 5}b_{3,2}(\mu) \left({ \alpha(Q^2) \over \alpha(\mu^2)
}\right)^{1.284}; \ \ \ \ \ \
{\cal M}_4 [{\tilde h}_L(Q^2)] ={1 \over 3}b_{4,2}(\mu)
\left({ \alpha(Q^2) \over \alpha(\mu^2) }\right)^{1.357}.
\end{eqnarray}
These curves normalized at $\mu=1$ GeV
are shown in Fig. 5.
Here and below we set
$N_f=3$ and $\Lambda_{QCD} =0.5$ GeV.
For comparison, we also plotted the
moments of the twist-2 distributions $f_1$
and $h_1$:
\begin{eqnarray} { {\cal M}_3 [f_1(Q^2)]
\over {\cal M}_3 [f_1(\mu^2)] }
=\left({ \alpha(Q^2) \over \alpha(\mu^2) }\right)^{0.775};
\ \ \ \ \ \ \ {
{\cal M}_3 [h_1(Q^2)] \over {\cal M}_3
[h_1(\mu^2)] } =\left({ \alpha(Q^2)
\over \alpha(\mu^2) }\right)^{0.790}.
\end{eqnarray}
{}From this figure, one can clearly
see that the third moment of the
twist-3 distribution evolves significantly
faster than that of the twist-2 structure function.

For $n=5$, the anomalous dimension becomes the $2 \times 2$
matrix:
\begin{eqnarray} X =
\left(\matrix{ -{202 \over 15}& {191 \over 90}\cr
{71 \over 60}&
-{3149 \over 180}\cr}\right) ,
\end{eqnarray}
and the eigenvalues of $\tilde{\gamma}\equiv
-X/(11-{2 \over 3} N_f)$
are 1.435 and 2.005.  We
thus get for ${\cal M}_5[{\tilde h}_L(Q^2)]$ as
\begin{eqnarray} {\cal M}_5 [{\tilde h}_L(Q^2)] &=& \left(
0.416 b_{5,2}(\mu) + 0.193b_{5,3}(\mu) \right)
\left({ \alpha(Q^2) \over
\alpha(\mu^2) }\right)^{1.435} \nonumber\\ &+&
\left( 0.013 b_{5,2}(\mu) -
0.050b_{5,3}(\mu) \right) \left({ \alpha(Q^2)
\over \alpha(\mu^2)
}\right)^{2.005} .
\end{eqnarray}
In principle, if one measures
${\cal M}_5[{\tilde h}_L(Q^2)]$ at
two different values of $Q^2$ with sufficient accuracy,
one could fix the two
matrix elements $b_{5,2}(\mu)$ and $b_{5,3}(\mu)$, and
the measurement of ${\cal
M}_5[{\tilde h}_L(Q^2)]$ at different $Q^2$ supplies a test of
the QCD evolution.
Since we do not have any physical insight on these matrix
elements, we plotted
${\cal M}_5 [{\tilde h}_L(Q^2)]$ normalized at $\mu = 1$
GeV in Fig. 6 with four
moderate values of $\lambda(\mu)=b_{5,3}(\mu)/b_{5,2}
(\mu)=-4.0,-2.0,1.0,4.0$ at
$\mu=1$ GeV.  One can see from the figure that there is a
large variation in the $Q^2$-evolution among
different choices of $\lambda(\mu)$.  This fact suggests
that a nucleon model
and a nonperturbative technique of QCD can be tested by
comparison of their
prediction on $\lambda(\mu)$ with future experiments.

As a measure of the large $Q^2$-behavior
of the moments, we have plotted in Fig. 7 the
lowest eigenvalues of the matrix
$\tilde{\gamma}_{lm}$ ($l,m = 2,.., [{n+1
\over 2}]$) with
$N_f=3$ for ${\tilde h}_L$, ${\tilde g}_T$ (defined as
the twist-3 part of $g_T$), $h_1$, and $f_1$ as
a function of $n$.  For
${\tilde g}_T$, we used the result for $\tilde{\gamma}$
obtained in
ref.\,\cite{JC}.  (The
authors of ref.\cite{JC} calculated
the anomalous dimension matrix of twist-3 operators
only for the
even moments of $g_T$, since they
discussed it in the context of the deep
inelastic scattering.)  From this figure we
expect that the moment of the
twist-3 distributions evolves faster than that of the
twist-2 distributions.  This fact
indicates that the measurement of $g_T$ and $h_L$
greatly serves as a new test
of QCD at high energies.  If one looks into more
detail of fig. 7 one sees that
the chiral-odd structure function, $h_1$ and ${\tilde h}_L$,
evolves slightly faster than
the chiral-even ones with the same twist,
$f_1$, $g_1$ and ${\tilde g}_T$.
In general,
however, actual form of the $Q^2$-evolution in a finite
 $Q^2$ window strongly
depends on the relative magnitude among $b_{n,l}(\mu)$
as we saw in the above
example for $n=5$.  We thus should take the result shown
in Fig. 7
only as a
rough measure for the asymptotic behavior
of the $Q^2$-evolution.

\section{Summary and Conclusion}
\setcounter{equation}{0}
\renewcommand{\theequation}{\arabic{section}.\arabic{equation}}

In this paper we have studied the $Q^2$-evolution of
the moments of the
chiral-odd twist-3 spin structure function $h_L(x,Q^2)$
in the standard QCD
perturbation theory.  For the $n$-th moment
of the twist-3 part of $h_L(x, Q^{2})$,
${\cal M}_n [{\tilde h}_L(Q^2)]$,
$\left[ {n+3 \over 2} \right]$
independent twist-3 operators, $R_{n,l}$ $(l=2,
\cdot\cdot\cdot,\left[
{n+1 \over 2} \right])$, $N_n$ and $E_n$ play roles.
The $Q^2$-evolution of ${\cal
M}_n [{\tilde h}_L(Q^2)]$ is governed by the anomalous
dimension matrix for the operators
$\{ R_{n,l}, N_n, E_n \}$.  We thus have calculated
the one-loop correction
to the three-point Green function which imbeds these
operators.
Although the physical (on-shell)
matrix element of the EOM (equation of motion)
operator $E_n$ vanishes,
$E_{n}$ mixes with $R_{n,l}$ and $N_{n}$ through
renormalization
and it is essential to take into account this operator
mixing to determine the anomalous dimension matrix.
In order to
incorporate the mixing correctly, we employed the off-shell
kinematics for the external lines.   Using the minimal
subtraction (MS) scheme
we obtained
the renormalization constants in the one-loop order
in the form of the
upper triangular matrix which really shows the mixing
among $\{ R_{n,l}, N_n,
E_n \}$.  As an example of the $Q^2$-evolution of $h_L$,
we have studied the
3-rd, 4-th and 5-th moments of $h_L(x,Q^2)$
in detail for $m_q=0$,
comparing
them with the known twist-2 distributions, $f_1$, $g_1$
and $h_1$,
and the other
twist-3 distribution $g_T$.
To consider the case $m_q=0$ is a sufficiently good
approximation for
the $u$ and $d$-quark
distributions, and is a reasonable one for the $s$-quark
distribution,
since the nucleon matrix element of $N_n$ is expected to be
small for the $s$-quark.
The notable features of the $Q^2$-evolution of these moments
can be summarized as follows:
\vskip 10pt
\begin{enumerate}

\item The $Q^2$-evolution of the 3-rd and 4-th moments
can be predicted
uniquely,
since there is only one relevant operator $R_{n,2}$
$(n=3,4)$.
(Note the
nucleon matrix element of $E_n$ vanishes.)
Compared with the twist-2 distributions
$f_1$, $g_1$ and $h_1$, the moments of ${\tilde h}_L(x,Q^2)$
evolves significantly faster.

\item The fifth or higher moments of ${\tilde h}_L$
receives contribution from two or more operators, and thus the
$Q^2$-evolution
 depends on the ratio among these matrix elements at
 a reference scale $\mu$.
They can, in principle, be determined by measuring
$h_L(x,Q^2)$ at several
values of
$Q^2$, and the measurement at different values
of $Q^2$ gives a test
for the QCD evolution.

\item The lowest eigenvalues of the anomalous
dimension matrix for
each moment of ${\tilde h}_L$ and ${\tilde g}_T$
(the twist-3 part of $g_T$)
are much larger than those of twist-2 distributions,
which
implies that
the $Q^2$-evolution of these twist-3 distributions
is significantly faster
than the twist-2 distributions.
These numbers for ${\tilde h}_L$
are slightly larger than those for ${\tilde g}_T$,
which suggests
that the chiral-odd distribution ${\tilde h}_L$
evolves faster than the chiral-even
one ${\tilde g}_T$.  The similar tendency has been
known for
the twist-2 distributions, $h_1$ and $g_1$.

\end{enumerate}

The anomalous dimension matrix for the twist-3
operators obtained in this paper
determines only the $Q^2$-evolution
of the moments of ${\tilde h}_L(x,Q^2)$.
In order to predict the $Q^2$-evolution of the
whole $x$-dependent
distribution ${\tilde h}_L(x,Q^2)$,
we need to construct a generalized
Altarelli-Parisi equation for a relevant
multi-parton distribution function.
This is because
a higher-twist distribution
essentially represents a correlated quark-gluon
distribution
and ${\tilde h}_L(x,Q^2)$ is only a
particular projection of this generalized
multi-parton distribution.  This work is under way
and will be published
in a future publication.
Nevertheless, we already found in this work some peculiar
features in the $Q^2$-evolution of the moments of
$h_L(x,Q^2)$,
which we hope will be measured in the future
collider experiments.

\vskip 20pt
\centerline{\bf Acknowledgement}
The authors
would like to thank J.Kodaira and T.Uematsu
for useful discussions, and T.Cheon for
supplying a computer code for the matrix diagonalization.
One of the authors (Y.K.) thanks X.Ji and C.-P.Yuan
for useful conversations.
The work of Y.K. at MSU was supported in part by the
US National Science
Foundation under grant PHY-9017077.
The work of K.T. was performed in part under the auspices of
Special Researchers' Basic Science Program of The Institute
of Physical and Chemical Research (RIKEN).

\newpage
\appendix
{\LARGE\bf Appendices}
\section*{A. One-loop Feynman amplitudes}
\setcounter{section}{1}
\setcounter{equation}{0}
\renewcommand{\theequation}{\Alph{section}.\arabic{equation}}

In this Appendix we present the expressions
for the one-loop Feynman
amplitudes for the truncated one-particle-irreducible
Green functions
with the insertion of the composite operators
${\cal O}_{i} = R_{n,l}\cdot \Delta$ ($l = 2, \cdots [
(n+1)/2]$), $E_{n}\cdot\Delta$ and $N_{n}\cdot\Delta$.

First we consider the three-point function with the insertion
of $R_{n,l} \cdot \Delta$.
We set $m_{q}=0$ following the discussion of (2) in sect.
\ref{sec3n}.
The Feynman diagrams for the one-loop corrections
are shown in Fig. 2.
The Feynman amplitudes are contracted by $\Omega_{\mu}$ and
are grouped to possess a definite charge conjugation.
(Note that, in the Feynman diagrams of Fig. 2,
(d), (f), and (h) are connected by the charge conjugation
to (c), (e), and (g),
respectively.)
Each group can be expanded
by the basic vertices ${\cal R}^{(3)}_{n,m}$ and
${\cal E}^{(3)}_{n}$.
(Here and in the following, we denote
${\cal R}_{n,m}^{(3)}
\cdot\Omega$
and ${\cal E}_{n}^{(3)}\cdot\Omega$ of (\ref{eq304}) and
(\ref{eq305}) simply by
${\cal R}_{n,m}^{(3)}$ and ${\cal E}_{n}^{(3)}$.
A sample calculation
of the diagrams (e) and (f) are given in Appendix B.)

Fig.2(a) gives 0.

Fig.2(b) gives:
\noindent
\begin{eqnarray}
\lefteqn{\frac{g^{2}}{16\pi^{2}\varepsilon}
C_{G} \left[ \sum_{m = 2}^{l-1} \left(\frac{(l+m)(m-1)}
{2 (l-m)[l-1]_{2}}
- \frac{(n-l+2+m)(m-1)}{2(n-l+2-m)[n-l+1]_{2}} \right)
{\cal R}^{(3)}_{n,m} \right.} \nonumber \\
& & + \left\{1 - S_{l} - S_{n-l+2}
+ \frac{1}{2 l} + \frac{1}{2(n-l+2)}
- \frac{(n+2)(l-1)}{2(n-2l+2)[n-l+1]_{2}}\right\}
{\cal R}^{(3)}_{n,l}
\nonumber \\
& &+ \left. \sum_{m = l+1}^{[(n+1)/2]}\left(
\frac{(2n-l-m+4)(n-m+1)}{2(m-l)[n-l+1]_{2}}
- \frac{(n-l+2+m)(m-1)}{2(n-l+2-m)[n-l+1]_{2}}\right)
{\cal R}^{(3)}_{n,m}
\right]. \label{eq:a4}
\end{eqnarray}

Fig.2(c) + Fig.2(d) gives:

\begin{eqnarray}
\lefteqn{\frac{g^{2}}{16\pi^{2}\varepsilon}(2 C_{F}-C_{G})
\left[
\sum_{m=2}^{l-1}2(-1)^{m}\left( \frac{_{n-l+1}C_{m-1}}
{[n-l+1]_{3}}
- \frac{_{l-1}C_{m-1}}{[l-1]_{3}}\right)
{\cal R}_{n,m}^{(3)}\right.}
\nonumber \\
& &+2(-1)^{l} \left( \frac{_{n-l+1}
C_{l-1} - (-1)^{n}}{[n-l+1]_{3}}
- \frac{1}{[l-1]_{3}} \right) {\cal R}_{n,l}^{(3)}
\nonumber\\
& &+ \sum_{m = l+1}^{[(n+1)/2]}
\frac{2(-1)^{m}}{[n-l+1]_{3}}
\left(_{n-l+1}C_{m-1} - (-1)^{n}\: _{n-l+1}C_{m-l}\right)
{\cal R}^{(3)}_{n,m} \nonumber \\
& & +\left. \left(\frac{1}{[l]_{2}} -
\frac{1}{[n-l+2]_{2}}\right)
{\cal E}^{(3)}_{n} \right]. \label{eq:a1}
\end{eqnarray}

Fig.2(e) + Fig.2(f) gives:

\begin{eqnarray}
\lefteqn{\frac{g^{2}}{16\pi^{2}\varepsilon}\left[
\sum_{m = 2}^{l-1}\left\{ (2C_{F}-C_{G})
\left(\frac{(-1)^{l-m}\:
_{l-2}C_{l-m}}{(l-m)\: _{n-m+1}C_{l-m}} -
\frac{(-1)^{n-l-m}\: _{n-l}C_{m-2}}{(n-l+2-m)
\: _{n-m+1}C_{l-1}}\right)
\right. \right.} \nonumber \\
& &- \left. C_{G}\left( \frac{1}{l} - \frac{1}{n-l+2}
\right)\right\}
{\cal R}_{n,m}^{(3)}
\nonumber \\
& &- \left\{ (2C_{F}-C_{G}) \frac{(-1)^{n}(l-1)}
{(n-2l+2)(n-l+1)}
+2C_{F}(S_{l-1} + S_{n-l+1} - 2) + C_{G}\frac{1}{l}
\right\} {\cal R}_{n,l}^{(3)} \nonumber \\
& &+ \sum_{m = l+1}^{[(n+1)/2]} (2C_{F}- C_{G}) \left(
\frac{(-1)^{m-l}\: _{n-l}C_{m-l}}{(m-l)\: _{m-1}C_{m-l}}
- \frac{(-1)^{n-l-m}\: _{n-l}C_{m-2}}
{(n-l+2-m) _{n-m+1}C_{l-1}}
\right) {\cal R}_{n,m}^{(3)} \nonumber \\
& & \left.+ C_{G}\left(\frac{1}{l} - \frac{1}{n-l+2}
\right){\cal E}_{n}^{(3)} \right].
\label{eq:a2}
\end{eqnarray}

Fig.2(g) + Fig.2(h) gives:

\begin{eqnarray}
\lefteqn{\frac{g^{2}}{16\pi^{2}\varepsilon}
C_{G}\left[\sum_{m=2}^{l-1} \frac{1}{2}\left(
\frac{(l-3)(l+1-m)}{[l-1]_{3}} + \frac{l+2}{[l-1]_{2}}
\right.\right.} \nonumber \\
& &-\left.\frac{(n-l-1)(n-l+3-m)}{[n-l+1]_{3}}
- \frac{n-l+4}{[n-l+1]_2}
\right) {\cal R}_{n,m}^{(3)}
\nonumber \\
& & + \frac{1}{2}\left( \frac{l-3}{[l-1]_{3}}
+ \frac{l+2}{[l-1]_{2}}
-\frac{(n-l-1)(n-2l+2)}{[n-l+1]_{3}}\right)
{\cal R}^{(3)}_{n,l}
\nonumber \\
& & - \sum_{m=l+1}^{[(n+1)/2]}\frac{(n-l-1)(n-2m+2)}
{2[n-l+1]_{3}}
{\cal R}^{(3)}_{n,m} \nonumber \\
& & - \left.\left(\frac{1}{l+1} - \frac{1}{n-l+3}
\right) {\cal E}^{(3)}_{n}
\right].
\label{eq:a3}
\end{eqnarray}

The coefficients in these expansions diverge as
$\varepsilon = (4-d)/2 \rightarrow 0$ with $d$
the space-time dimension.
By adding the counter term contribution (\ref{eq302n})
to the sum of eqs.(\ref{eq:a4})-(\ref{eq:a3})
and by requiring that those counter term
contributions cancel the
$1/\varepsilon$ pole terms,
we obtain the renormalization constants given by
(\ref{eq313})-(\ref{eq316}) in the MS scheme.

Next we proceed to calculate the two-point
function with the
insertion of $R_{n,l}\cdot \Delta$,
$E_{n}\cdot\Delta$ and
$N_{n}\cdot\Delta$, following the discussion of
(3) in sect. \ref{sec3n}.
We keep the quark mass $m_{q}$ as a nonzero quantity.
The computation can be performed in a similar
manner as in the case of
the three-point functions. The three-point
vertices necessary to
compute the one-loop diagrams of Fig. 3 are given by
(\ref{eq301})-(\ref{eq303n}).
Again the amplitudes combined appropriately
can be expanded by the basic vertices (\ref{eq309}) and
(\ref{eq310}).

The one-loop correction to the two-point function
with $R_{n,l}\cdot\Delta$ comes from
Figs.3 (b) and (c).  It gives
\begin{eqnarray}
\lefteqn{\frac{g^{2}}{16 \pi^{2} \varepsilon}C_{F}
\left[ \left(\frac{2}{[l]_{2}}-
\frac{2}{[n-l+2]_{2}}\right)
{\cal E}^{(2)}\right.}
\nonumber\\
& &- \left.\left(\frac{4}{[l-1]_{3}}-\frac{4}
{[n-l+1]_{3}}\right)
{\cal N}^{(2)} \right].
\label{eq:a10}
\end{eqnarray}

For the one-loop correction to the one with
$E_{n}\cdot\Delta$,
Fig.3(a) gives
\begin{equation}
\frac{g^{2}}{16\pi^{2}\varepsilon}C_{F}
\frac{2}{n}{\cal N}^{(2)},
\label{eq:a11}
\end{equation}
and Fig.3(b)+Fig.3(c) gives
\begin{equation}
\frac{g^{2}}{16\pi^{2}\varepsilon}C_{F}
\left[ \left(1 - 2 \sum_{j=2}^{n}\frac{1}{j}
\right){\cal E}^{(2)}
- \left(3 + \frac{2}{n}\right){\cal N}^{(2)}\right].
\label{eq:a12}
\end{equation}

Finally, for the one-loop correction to
the one with $N_{n}\cdot\Delta$,
Fig.3(a) gives
\begin{equation}
\frac{g^{2}}{16\pi^{2}\varepsilon} C_{F} \frac{2}{n(n+1)}
{\cal N}^{(2)},
\label{eq:a13}
\end{equation}
while each of Figs.3 (b) and (c) gives
the same contribution,
\begin{equation}
\frac{g^{2}}{16\pi^{2}\varepsilon}C_{F}\left(- 2
\sum_{j=2}^{n}
\frac{1}{j}\right) {\cal N}^{(2)}.
\label{eq:a14}
\end{equation}

By adding the appropriate counter term contributions,
we can determine the counter terms in the MS scheme,
which give the results (\ref{eq316})-(\ref{eq320}).
Note that (\ref{eq316}) are obtained from the
three-point functions
as well as from the
two-point functions,
giving a consistency check of our methods.
Also, $Z_{EN}=0$ can be verified explicitly
by using the results
(\ref{eq:a11}) and (\ref{eq:a12}) (see the discussion
of (3) in
sect. \ref{sec3n}).

\section*{B. Sample calculation}
\setcounter{section}{2}
\setcounter{equation}{0}
\renewcommand{\theequation}{\Alph{section}.\arabic{equation}}

In this appendix we describe the details of the calculation
of the one-loop Feynman amplitudes.
We choose the diagram (e) and (f) of Fig.2
with the insertion of
$R_{n,l}\cdot \Delta$ as an example; the other diagrams
can be
calculated using the similar technique.

First we write down the Feynman amplitude ${\cal F}_{(e)}$
for the diagram (e) by using the vertex (\ref{eq302}) and
the usual Feynman rule (in the Feynman gauge):
\begin{eqnarray}
\lefteqn{{\cal F}_{(e)}=
\mu^{2 \varepsilon} \int
\frac{{\rm d}^{4 - 2\varepsilon}k}{(2\pi)^{4-2\varepsilon}}
igt^{b}\gamma_{\nu}
\frac{i\left(\rlap/{\mkern-1mu p} + \rlap/{\mkern-1mu
k}\right)}
{\left(p - k\right)^{2}}
\left(- \frac{g^{2}}{2} \Omega_{\mu}
\sigma^{\mu \lambda} \Delta_{\lambda} i \gamma_{5}
\Delta_{\rho}
\right)}
\nonumber \\
& & \times
\left[ \left\{(\hat{p} - \hat{k})^{n-l}\hat{q}^{l-2}
- (\hat{p} - \hat{k})^{l-2}\hat{q}^{n-l} \right\}
if^{abc}t^{c}
\right.
\nonumber \\
& & + \left\{ \sum_{j=2}^{n-l+1}(\hat{p}-\hat{k})^{j-2}
\hat{p}^{n-l+1-j} \hat{q}^{l-2}
- \sum_{j=2}^{l-1}(\hat{p} - \hat{k})^{j-2}
\hat{p}^{l-1-j}\hat{q}^{n-l} \right\}(\hat{q}-\hat{p})
t^{b}t^{c}
\nonumber \\
& & +
\left\{ \sum_{j=n-l+3}^{n} (\hat{p}-\hat{k})^{n-l}
(\hat{q}-\hat{k})^{j-3-n+l} \hat{q}^{n-j}\right.
\nonumber \\
& &- \left.\left.\sum_{j=l+1}^{n}(\hat{p}-\hat{k})^{l-2}
(\hat{q}-\hat{k})^{j-1-l}
\hat{q}^{n-j} \right\} (\hat{q}-\hat{p}) t^{a}t^{b} \right]
\frac{-i g^{\nu \rho}}{k^{2}},
\label{eq:a}
\end{eqnarray}
where $q$ and $p$ are the incoming and the outgoing
off-shell quark momenta, and $k$ is the gluon loop momentum.
We work in the massless quark limit ($m_q=0$)
following (2) of sect. 3.
The Lorentz index $\mu$ corresponding to the external
gluon line
is contracted by $\Omega_{\mu}$,
which kills off the terms involving $\Delta_{\mu}$
in the vertex (\ref{eq302}). Also, owing to this property,
the factors involving the gamma matrices
can be easily evaluated:
\begin{equation}
\gamma_{\nu} (\rlap/{\mkern-1mu p}
+ \rlap/{\mkern-1mu k}) \Omega_{\mu}
\sigma^{\mu \lambda} \Delta_{\lambda}i\gamma_{5}
\Delta_{\rho}
g^{\nu \rho}
= 2 (\hat{p} + \hat{k}) \Omega_{\mu}
\sigma^{\mu \lambda} \Delta_{\lambda}i\gamma_{5}.
\label{eq:i}
\end{equation}
We follow the standard procedure:
We use the Feynman parameterization to collect
all the denominators of the quark and the gluon propagators.
After shifting the $k$-integration,
many terms can be dropped by the condition $\Delta^{2}=0$.
We retain only the divergent
contribution of the $k$-integration:
\begin{eqnarray}
\lefteqn{{\cal F}_{(e)} =
g^{3} \frac{1}{16\pi^{2}\varepsilon} \Omega_{\mu}
\sigma^{\mu \lambda} \Delta_{\lambda} i\gamma_{5}t^{a}
\int_{0}^{1} {\rm d} x}
\nonumber \\
& & \times
\left\{ - \frac{C_{G}}{2} \left( \hat{t}^{n-l+1}
\hat{q}^{l-2}
- \hat{t}^{l-1} \hat{q}^{n-l} \right)\right.
\nonumber \\
& & + C_{F}\left(\sum_{j=2}^{n-l+1} \hat{t}^{j-1}
\hat{p}^{n-l+1-j}
\hat{q}^{l-2} - \sum_{j=2}^{l-1} \hat{t}^{j-1}\hat{p}^{l-1-j}
\hat{q}^{n-l}\right) (\hat{q}-\hat{p})
\nonumber \\
& & + \left.\left(C_{F} - \frac{C_{G}}{2} \right)
\left( \sum_{j=n-l+3}^{n} \hat{t}^{n-l+1} \hat{s}^{j-3-n+l}
\hat{q}^{n-j} - \sum_{j=l+1}^{n} \hat{t}^{l-1}
\hat{s}^{j-1-l}
\hat{q}^{n-j} \right) (\hat{q}-\hat{p}) \right\},
\label{eq:u}
\end{eqnarray}
where $t=(1-x)p$, $s=-xp + q$.
We perform the Feynman parameter integral by
\begin{equation}
\int_{0}^{1} {\rm d}x \hat{t}^{l-1}\hat{s}^{j-1-l}
= \frac{1}{l}\sum_{r=0}^{j-1-l} (-1)^{r}
\frac{_{j-1-l}C_{r}}
{_{l+r}C_{r}} \hat{p}^{l-1+r}\hat{q}^{j-1-l-r}.
\label{eq:o}
\end{equation}
The resulting formula of ${\cal F}_{(e)}$
involves double summation.
These terms can be simplified into single summation
by interchanging the order of summation and by using
the relation $\sum_{r=m}^{n}\: _{r}C_{m} = \:_{n+1}C_{m+1}$.
Then, we obtain
\begin{eqnarray}
{\cal F}_{(e)} & &
= 2 \frac{g^{2}}{16\pi^{2}\varepsilon}
\frac{g}{2}\Omega_{\mu}\sigma^{\mu \lambda}
\Delta_{\lambda}i\gamma_{5}t^{a}
\left[ \frac{C_{G}}{2}\left(\frac{1}{l}
\hat{p}^{l-1}\hat{q}^{n-l}
- \frac{1}{n-l+2} \hat{p}^{n-l+1} \hat{q}^{l-2}
\right)\right.
\nonumber \\
& & + C_{F} \left\{ \left(S_{n-l+1} -1\right)
\hat{p}^{n-l}\hat{q}^{l-2} - \left(S_{l-1}-1\right)
\hat{p}^{l-2}\hat{q}^{n-l} \right\}(\hat{q}-\hat{p})
\nonumber \\
& & + \left(C_{F} - \frac{C_{G}}{2} \right)
\left\{ \sum_{m=2}^{l-1} \frac{(-1)^{l-m}\:_{l-2}C_{l-m}}
{(l-m)\:_{n-m+1}C_{l-m}} \hat{p}^{n-m}\hat{q}^{m-2}
\right.
\nonumber \\
& &-\left. \left.\sum_{m=l+1}^{n}\frac{(-1)^{m-l}
\: _{n-l}C_{m-l}}
{(m-l) \:_{m-1}C_{m-l}} \hat{p}^{m-2}\hat{q}^{n-m}
\right\} (\hat{q}-\hat{p})\right].
\label{eq:ki}
\end{eqnarray}

The amplitude corresponding to the diagram (f)
can be computed in a similar manner as above.
The result should be related to (\ref{eq:ki})
by charge conjugation.
By adding this result to (\ref{eq:ki}),
we obtain as the total result:
\begin{eqnarray}
{\cal F} & &\equiv {\cal F}_{(e)}+ {\cal F}_{(f)}
\nonumber \\
& & = 2 \frac{g^{2}}{16\pi^{2}\varepsilon}
\frac{g}{2} \Omega_{\mu}\sigma^{\mu \lambda}
\Delta_{\lambda}
i \gamma_{5} t^{a}
\nonumber \\
& & \times \left[ \frac{C_{G}}{2} \left\{
\frac{1}{l}
\left( \hat{p}^{n-l}\hat{q}^{l-1}+ \hat{p}^{l-1}
\hat{q}^{n-l}\right)
- \frac{1}{n-l+2}\left(\hat{p}^{n-l+1}\hat{q}^{l-2}
+ \hat{p}^{l-2}
\hat{q}^{n-l+1} \right) \right\}\right.
\nonumber \\
& &+C_{F}(S_{l-1} + S_{n-l+1} -2) \left(\hat{p}^{n-l}
\hat{q}^{l-2}
- \hat{p}^{l-2}\hat{q}^{n-l} \right) (\hat{q}-\hat{p})
\nonumber \\
& & - \left( C_{F}-\frac{C_{G}}{2}\right)
\left\{ \sum_{m=2}^{l-1}\frac{(-1)^{l-m}\:_{l-2}C_{l-m}}
{(l-m)\:_{n-m+1}C_{l-m}}\left(\hat{p}^{n-m}\hat{q}^{m-2}
-\hat{p}^{m-2}\hat{q}^{n-m}\right)(\hat{q}-\hat{p}) \right.
\nonumber \\
& & + \left. \left.\sum_{m=l+1}^{n}\frac{(-1)^{m-l}
\:_{n-l}C_{m-l}}
{(m-l)\:_{m-1}C_{m-l}} \left(\hat{p}^{n-m}\hat{q}^{m-2}
-\hat{p}^{m-2}\hat{q}^{n-m} \right)(\hat{q}-\hat{p})
\right\} \right].
\label{eq:ku}
\end{eqnarray}
This should be expressed as a linear combination
of the basic vertices
(\ref{eq304}) and (\ref{eq305}), which we denote simply by
${\cal R}_{n,l}^{(3)}$ and ${\cal E}_{n}^{(3)}$.
This is readily performed for the terms proportional to
$C_{F}$ or $C_{F}-C_{G}/2$. For the terms
proportional to $C_{G}$,
we use the identity
\begin{eqnarray}
\lefteqn{\hat{p}^{n-l}\hat{q}^{l-1}
+ \hat{p}^{l-1}\hat{q}^{n-l}}
\nonumber \\
& &= \hat{p}^{n-1} +\hat{q}^{n-1}
+\sum_{m=2}^{l}\left(\hat{p}^{n-m}\hat{q}^{m-2}-\hat{p}^{m-2}
\hat{q}^{n-m} \right) (\hat{q}-\hat{p})
\nonumber \\
& & = \hat{p}^{n-1} +\hat{q}^{n-1}
- \sum_{m=l+1}^{n}\left(\hat{p}^{n-m}\hat{q}^{m-2}
- \hat{p}^{m-2}\hat{q}^{n-m} \right)(\hat{q}-\hat{p}).
\label{eq:ke}
\end{eqnarray}
Now we obtain
\begin{eqnarray}
\lefteqn{{\cal F}= -2 \frac{g^{2}}{16\pi^{2}
\varepsilon}\left\{
C_{F}\left(S_{l-1}+S_{n-l+1}-2\right){\cal R}^{(3)}_{n,l}
+\frac{C_{G}}{2}\left(\frac{1}{l} \sum_{m=2}^{l}
{\cal R}^{(3)}_{n,m}
+ \frac{1}{n-l+2}\sum_{m=l}^{n}{\cal R}^{(3)}_{n,m}
\right)\right.}
\nonumber \\
& &- \left(C_{F}- \frac{C_{G}}{2}\right)\left(
\sum_{m=2}^{l-1}\frac{(-1)^{l-m}\:_{l-2}C_{l-m}}
{(l-m)\:_{n-m+1}
C_{l-m}} {\cal R}^{(3)}_{n,m} +
\sum_{m=l+1}^{n} \frac{(-1)^{m-l}\:_{n-l}C_{m-l}}
{(m-l)\:_{m-1}
C_{m-l}}  {\cal R}^{(3)}_{n,m} \right)
\nonumber \\
& &\left. + \frac{C_{G}}{2} \left(\frac{1}{n-l+2}
- \frac{1}{l}
\right) {\cal E}_{n}^{(3)}\right\}.
\label{eq:ko}
\end{eqnarray}
Here ${\cal R}^{(3)}_{n,m}$ with $m = [(n+1)/2] + 1,
\cdots, n$ appear.
These can be expressed by those for
$m = 2, \cdots, [(n+1)/2]$ using
${\cal R}^{(3)}_{n,n-m+2} = - {\cal R}^{(3)}_{n,m}$,
and we obtain (\ref{eq:a2}) of Appendix A.

\newpage
\centerline{\bf Figure Captions}

\vskip 15pt
\begin{description}

\item[Fig. 1] (a) Three-point basic vertex for $R_{n,l}$,
$E_n$ and
$N_n$.  (b) Four-point basic vertex for $R_{n,l}$
necessary for
the calculation
of the diagrams shown in Fig. 2.


\item[Fig. 2] One-particle-irreducible diagrams
for the one-loop correction
to the three-point Green function $F_i(p,q,k)$
(eq.(\ref{eq226})).


\item[Fig. 3]  One-loop corrections to the two-point
function
relevant for the
calculation of $Z_{lE}$, $Z_{lN}$.   These diagrams are
also used for the
calculation of the anomalous dimension of the twist-2
distributions.


\item[Fig. 4]  Diagrams which could cause mixing
between a flavor-singlet
quark distribution $h_1$ and a gluon distribution (if any).
These diagrams are identically zero for the chiral-odd
distribution.


\item[Fig. 5]  The $Q^2$-evolution of the 3-rd and 4-th
moments of
${\tilde h}_L(x,Q^2)$ normalized at $\mu = 1$ GeV.
The 3-rd moments of
twist-2 distributions $f_1$ and $h_1$ are also plotted
for comparison.


\item[Fig. 6]  The $Q^2$-evolution of the 5-th moment of
${\tilde h}_L(x,Q^2)$ normalized at $\mu = 1$ GeV for
four moderate values
of $\lambda(\mu) = -4.0, -2.0, 1.0, 4.0$.


\item[Fig. 7]
The smallest eigenvalues of ${\tilde \gamma}
= -X/16\pi^2 \beta_0$ as a function
of the dimension of the moment, $n$.

\end{description}


\newpage

\begin{thebibliography}{99}
\bibliographystyle{unsrt}
\setlength{\itemsep}{0.0in}

\bibitem{EMC} J. Ashman et al. (EMC
Collaboration),
Phys.\ Lett.\ {\bf B206} (1988)
364.\\
For more recent data, see
B. Adeva et al. (SMC
Collaboration), Phys.\ Lett. {\bf B302}
(1993) 533;\\
P.L. Anthony et al. (E142 Collaboration),
Phys.\ Rev.\ Lett.
{\bf 71} (1993) 959;\\
D. Adams et al. (SMC
Collaboration), Phys.\ Lett. {\bf B329}
(1994) 399.

\bibitem{AM} X. Artru and M. Mekhfi,
Z.Phys. {\bf C45} (1990) 669.

\bibitem{CPR} J.L. Cortes, B. Pire and J.P. Ralston,
Z.Phys. {\bf C55} (1992)
409.

\bibitem{JJ} R.L. Jaffe and X. Ji, Phys.\ Rev.\ Lett.\
{\bf 67} (1991) 552;
Nucl.\ Phys.\ {\bf B375} (1992) 527.

\bibitem{C} J.C. Collins, Nucl.\ Phys.\ {\bf B394}
(1993) 169.

\bibitem{RS} J. Ralston and D.E. Soper, Nucl.\ Phys.\
{\bf B152} (1979) 109.

\bibitem{JJ2} R.L. Jaffe, Comm. in Nucl. Part. Phys.
{\bf 19} (1990) 239.

\bibitem{ad} The first measurement of $g_{2}$ has been
reported in D. Adams et al. (SMC), Phys.\
Lett. {\bf B336} (1994) 125.

\bibitem{AP} G. Altarelli and G. Parisi, Nucl.\ Phys.\
{\bf B126} (1977) 298.


\bibitem{BKL} A.P.Bukhvostov,
E.A. Kuraev and L.N. Lipatov, Sov. Phys. JETP.,
{\bf 60}
(1984)  22.

\bibitem{Rat} P.G. Ratcliffe, Nucl.\ Phys.\ {\bf B264}
(1986) 493.

\bibitem{JC} X. Ji and C. Chou, Phys.\ Rev.\ {\bf D42}
(1990) 3637.

\bibitem{ABH} A. Ali, V.M. Braun and G. Hiller,
Phys.\ Lett.\ {\bf B266} (1991) 117.

\bibitem{SV} E. V. Shuryak and A. I. Vainshtein,
Nucl.\ Phys.\ {\bf B201} (1982) 141.

\bibitem{KU} J. Kodaira, Y. Yasui and T. Uematsu,
HUPD-9411, KUCP-71 (1994).

\bibitem{P80} H. D. Politzer,  Nucl.\ Phys.\
{\bf B172} (1980) 349.

\bibitem{JC2} J.C.Collins, {\it Renormalization}
(Cambridge Univ. Press, 1984).

\bibitem{CSS} For a general review of QCD
factorization see, for example,
J.C. Collins, D. Soper and G. Sterman,
Factorization of hard processes
in QCD, in {\it Perturbative Quantum
Chromodynamics (World Scientific,
Singapore, 1989, ed. A.H. Mueller)}.

\bibitem{QS2} For the factorization
of the higher twist,
see J. Qiu and G. Sterman, Nucl.\ Phys.\ {\bf B353}
(1991) 137

\bibitem{WW} W. Wandzura and F. Wilczek, Phys.\ Lett.\
{\bf B172} (1975) 195.

\bibitem{JS} R.L. Jaffe and M. Soldate,
Phys.\ Lett.\ {\bf B105} (1981) 467, Phys.\ Rev.\
  {\bf D26} (1982) 49.


\bibitem{Muta} T. Muta,
{\it Foundations of Quantum Chromodynamics}
(World Scientific, Singapore, 1987).




\end{thebibliography}
\end{document}